\documentclass[twocolumn,times,final, number]{elsarticle}

\usepackage{amssymb}

\usepackage{prletters}
\usepackage{framed,multirow}

\usepackage{amssymb}
\usepackage{latexsym}
\usepackage{dsfont}

\usepackage{url}
\usepackage{algorithm2e}
\usepackage{booktabs}
\usepackage{xcolor}
\definecolor{newcolor}{rgb}{.8,.349,.1}

\newcommand{\myRed}[1]{\textcolor{black}{#1}}

\journal{Pattern Recognition Letters}

\usepackage{amsmath}

\usepackage{soul}
\usepackage{pdfpages}

\usepackage[labelsep=endash]{caption}
\usepackage{subcaption}

\usepackage[export]{adjustbox}
\usepackage{gensymb}

\newcommand{\reftab}[1]{\autoref{#1}}
\newcommand{\citeN}[1]{}

\newcommand{\citeNN}[1]{}

\usepackage{glossaries}
\makeglossaries
\glsaddall

\usepackage[colorlinks=true]{hyperref}%
\usepackage[english]{babel}
\usepackage[figurename=Fig., tablename=Table.]{caption}

\addto\extrasenglish{%
}

\usepackage{adjustbox}
\usepackage{graphicx}
\usepackage[normalem]{ulem}
\useunder{\uline}{\ul}{}

\usepackage[para,online,flushleft]{threeparttable}

\usepackage{hyperref}
 \usepackage{float}
  \usepackage{footnote}
\usepackage{tablefootnote}

\definecolor{added}{RGB}{255,255,0}

\usepackage{float}
\usepackage{placeins}

\begin{document}

\begin{table*}[!t]

\section*{Research Highlights}
\vskip1pc
\fboxsep=6pt
\fbox{
\begin{minipage}{.95\textwidth}
Highlights should contain 3 to 5 bullet points with a maximum of 125 characters per bullet point, including spaces. Only the core results of the paper should be covered. Please ensure each bullet point is less than 125 characters
\vskip1pc
\begin{itemize}

 \item Encoder booster and bottleneck enhancement blocks are introduced to reduce feature loss in the encoder-decoder architecture
 \item An adaptive threshold algorithm is proposed to find the optimal threshold for pixelwise classification
 \item Micro-retinal vessel extraction was improved significantly
 \item A cross-training evaluation was performed between the DRIVE and STARE datasets
\end{itemize}
\vskip1pc
\end{minipage}
}

\end{table*}

\clearpage

\setcounter{page}{1}

\begin{frontmatter}

\title{FS-Net: Full Scale Network and Adaptive Threshold for Improving Extraction of Micro-Retinal Vessel Structures}

\author[inst1]{Melaku N. Getahun}
\ead{Melaku.Getahun@skoltech.ru}
\author[inst3,inst6,inst7]{Oleg Y. Rogov}
\ead{o.rogov@airi.net}
\author[inst1,inst3]{Dmitry V. Dylov}
\ead{d.dylov@skoltech.ru}
\author[inst1]{Andrey Somov}
\ead{a.somov@skoltech.ru}
\author[inst2,inst6]{Ahmed Bouridane}
\ead{abouridane@sharjah.ac.ae}
\author[inst4,inst5,inst6] {Rifat Hamoudi}
\ead{rhamoudi@sharjah.ac.ae}

\affiliation[inst1]{organization={Skolkovo Institute of Science and Technology (Skoltech)}, city={Moscow}, country={Russia}}
\affiliation[inst2]{organization={Center for Data Analytics and Cybersecurity, University of Sharjah}, city={Sharjah}, country={United Arab Emirates}}
\affiliation[inst3]{organization={Artificial Intelligence Research Institute (AIRI)}, city={Moscow}, country={Russia}}
\affiliation[inst4]{organization={Research Institute for Medical and Health Sciences, College of Medicine, University of Sharjah}, city={Sharjah}, country={United Arab Emirates}}
\affiliation[inst5] {organization={Division of Surgery and Interventional Science, Faculty of Medical Science, University College London}, city={London}, country={United Kingdom}}
\affiliation[inst6] {organization={BIMAI-Lab, Biomedically Informed Artificial Intelligence Laboratory, University of Sharjah}, city={Sharjah}, country={United Arab Emirates}}
\affiliation[inst7] {organization={Bian Que AI}, city={Hong Kong}, country={China}}

\received{10 September 2013}
\finalform{10 May 2013}
\accepted{13 May 2013}
\availableonline{15 May 2013}
\communicated{S. Sarkar}


\begin{abstract}
Retinal vascular segmentation, a widely researched topic in biomedical image processing, aims to reduce the workload of ophthalmologists in treating and detecting retinal disorders. Segmenting retinal vessels presents unique challenges; previous techniques often failed to effectively segment branches and microvascular structures. Recent neural network approaches struggle to balance local and global properties and frequently miss tiny end vessels, hindering the achievement of desired results. To address these issues in retinal vessel segmentation, we propose a comprehensive micro-vessel extraction mechanism based on an encoder-decoder neural network architecture. This network includes residual, encoder booster, bottleneck enhancement, squeeze, and excitation building blocks. These components synergistically enhance feature extraction and improve the prediction accuracy of the segmentation map. Our solution has been evaluated using the DRIVE, CHASE-DB1, and STARE datasets, yielding competitive results compared to previous studies. \myRed{The AUC and accuracy on the DRIVE dataset are 0.9884 and 0.9702, respectively. For the CHASE-DB1 dataset, these scores are 0.9903 and 0.9755, respectively, and for the STARE dataset, they are 0.9916 and 0.9750.} Given its accurate and robust performance, the proposed approach is a solid candidate for being implemented in real-life diagnostic centers and aiding ophthalmologists.
\end{abstract}

\begin{keyword}
 \sep Retinal Vascular Segmentation
 \sep Micorvascular Structures
 \sep Encoder Booster
 \sep Bottleneck Enhancement
 \sep Encoder-Decoder
\end{keyword}

\end{frontmatter}



\section{Introduction}
\label{sec:introduction}
Medical images are crucial for diagnoses \cite{cavaro2010diagnostic}, providing valuable information for medical experts to perform necessary procedures. However, these images can be noisy and difficult to interpret, particularly at the microlevel. This challenge is especially noticeable in retinal image data, where discerning vascular structures impacted by ophthalmological and cardiovascular diseases can be complex.

To diagnose and treat these diseases, the morphological attributes of the retinal vessels are fundamental, including length, width, tortuosity, branching patterns, and angles \cite{yao2016convolutional}. Although the doctors have the image obtained from medical equipment \cite{liu2019retinal}, it cannot show the exact structural elements due to imaging equipment limitations and the inherent characteristics of biological tissues. Eventually, the medical experts are required to make the segmentation manually, which is time-consuming and highly prone to bias.

For instance, the most common disease, diabetic retinopathy, which is caused by a higher complication in diabetics and leads to blindness, involves the thickening of the retinal blood vessels' wall, microvascular occlusions, retinal tissue ischemia, and finally the growth of abnormal new blood vessels, leading to severe complications and the loss of functional vision \cite{wang2009retinal}. 

Manual segmentation of retinal vessels complicates difficult situations further, making automatic retinal vessel segmentation a significant research area.

With the advent of high-performance computing devices and convolutional neural network technology, biomedical image segmentation has made significant improvements. In the case of specific architectures with symmetric contracting and expanding paths based on a fully convolutional network, called U-Net \cite{ronneberger2015u}, a high number of successful research works have been proposed with extension and modification in terms of their general structure. This fully convolutional network architecture has been used in different domains for segmentation tasks \cite{kazerouni2021ghost, xiao2020segmentation, ICARCV-NIR}. In addition, the current transformer networks are not able to reach their performance limit \cite{liu2021swin}. However, the plain U-Net architecture also has its own limitations, including segmentation of different scale inputs. For a better understanding, we would like to coin a classification of semantic segmentation into two types: small- and large-scale semantic segmentation.


Semantic segmentation involves focusing on each pixel in an image, with the first type focusing on finer details. Large-scale semantic segmentation involves large-scale objects visible to both neural network architectures and human naked eyes. Blood vessel segmentation is small-scale, with details and objects small enough to be invisible to the naked eye. U-Net architectures struggle to perform well due to retinal vessel sizes and microstructures.

To address these issues, several studies have been conducted, resulting in a number of architectures that have demonstrated improvements, including the salient U-Net \cite{hu2019s}, structured dropout U-Net \cite{guo2019sd}, deep residual U-Net \cite{liu2023resdo}, to name a few. Notwithstanding significant advancements in the extraction of vascular structures, many studies still suffer from microvascular segmentation. Hence, a methodology that can bridge this gap is much desired.

This paper proposes a novel full-scale neural network architecture based on the encoder-decoder topology that has a higher capability of extracting micro-vessels compared to previous works.
The contributions of this work are as follows:
\begin{enumerate}
    \item An encoder booster block is designed for minimizing the spatial loss of the microvascular structures that occur during feature extraction within the encoder block.
    \item A bottleneck enhancement module is proposed to enhance the features that are usually degraded before the up-sampler starts localizing the feature maps.
    \item An adaptive threshold algorithm is introduced in contrast to the common 0.5 threshold used for identifying the pixels as vessels and background after the sigmoid shift is applied to the output of the model's prediction.
\end{enumerate}

This article discusses retinal vessel segmentation studies such as CNN models, generative models, transformer models, and state-of-the-art works. It describes a proposed approach, the dataset used, and the results achieved, and completes with a concluding remark.


\section{Related Work}
\paragraph{Unsupervised Methods}
\myRed{This branch of machine learning is also applicable in scenarios such as image segmentation, as it has the advantage of not requiring labeled data. Unsupervised methods for image segmentation typically involve three key steps: pre-processing, segmentation, and post-processing. Upadhyay et al. \cite{upadhyay2020unsupervised} applied two multi-scale transformations—wavelets and curvelets. The former is used to enhance local structures, while the latter is employed to suppress non-vessel backgrounds. Tavakoli et al. \cite{tavakoli2021unsupervised} introduced the use of random transformations and overlapping windows to detect and locate a vascular tree within a fundus image. Alhussein et al. \cite{alhussein2020unsupervised} proposed the application of the Hessian matrix approach to extract both thick and thin vessels, as well as intensity transformation to maximize vessel details. Unsupervised learning relies on pixel colors and intensities while often overlooking contextual information and semantic understanding. In contrast, supervised learning methods such as deep learning yield better outcomes for image data, as they can learn complex features through hierarchical feature learning. This capability enables them to capture detailed information that might be missed by unsupervised methods, such as thin vessels.}

\paragraph{CNN Methods}
Retinal vessel segmentation aims to classify each pixel in a retinal image as either a vessel or background. U-Net \cite{ronneberger2015u} is commonly used for biomedical image segmentation. To prevent overfitting, methods like regularization and dropout layers are employed \cite{jindal2016learning}. Spatial information in convolutional layers can lead to overfitting, as noted in SD-U-Net \cite{guo2019sd}. A structured drop block method is suggested to mitigate this. Training neural networks with dynamic weights was used to overcome problems such as under-segmentation of faint vessels and edge pixels \cite{khanal2020dynamic}. 

Both supervised and unsupervised algorithms are combined for improved segmentation, incorporating multi-scale filters and U-Net models \cite{ma2021multichannel}. Edge detection and artificial neural networks help in identifying vessel pixels. Hard attention mechanisms with multiple decoder networks target different segmentation regions \cite{wang2020hard}. Single-step networks can misclassify pixels, leading to the development of NFN+ (Network after Network) with dual multi-scale backbones \cite{wu2020nfn+}. Multi-scale dense networks address vessel morphology complexity. Augmentation modules, like channel-wise random Gamma correction, are proposed to handle data variation \cite{sun2021robust}.

Study group learning (SGL) \cite{zhou2021study} proposed a method for addressing overfitting in small datasets by automatically erasing vessel segments in the vessel segmentation label. The current state of the art is also U-Net-based, called FR-UNet (Full resolution U-Net). The authors in \cite{liu2022full} claim that the previous state-of-the-art \cite{kamran2021rv} is characterized by lower sensitivity metrics, implying that the architecture is not capable of extracting thin vessels efficiently. FR-UNet includes a feature aggregation module that is embedded before each convolution block to aggregate feature maps from up-sampling and down-sampling. This helps extract multi-scale, high-level contextual information. FR-UNet has achieved the highest sensitivity of 0.8316, but there is room for improvement in capturing micro-vessel structures.

Despite advancements, limitations exist in detecting thin vessels and in small-scale semantic segmentation. Subsequent methods, while better than traditional U-Net, have not significantly improved branch vessel segmentation.

\paragraph{Generative-based methods} 
Kamran et al. \cite{kamran2021rv} developed the RV-GAN architecture, a GAN-based model for microvascular localization and segmentation. Despite excellent accuracy, the model's low sensitivity score suggests it struggles with extracting and localizing thin and branching vascular structures.

\begin{figure*}[ht]
\centering
\begin{subfigure}{.5\textwidth}
  \centering
  \includegraphics[width=\linewidth]{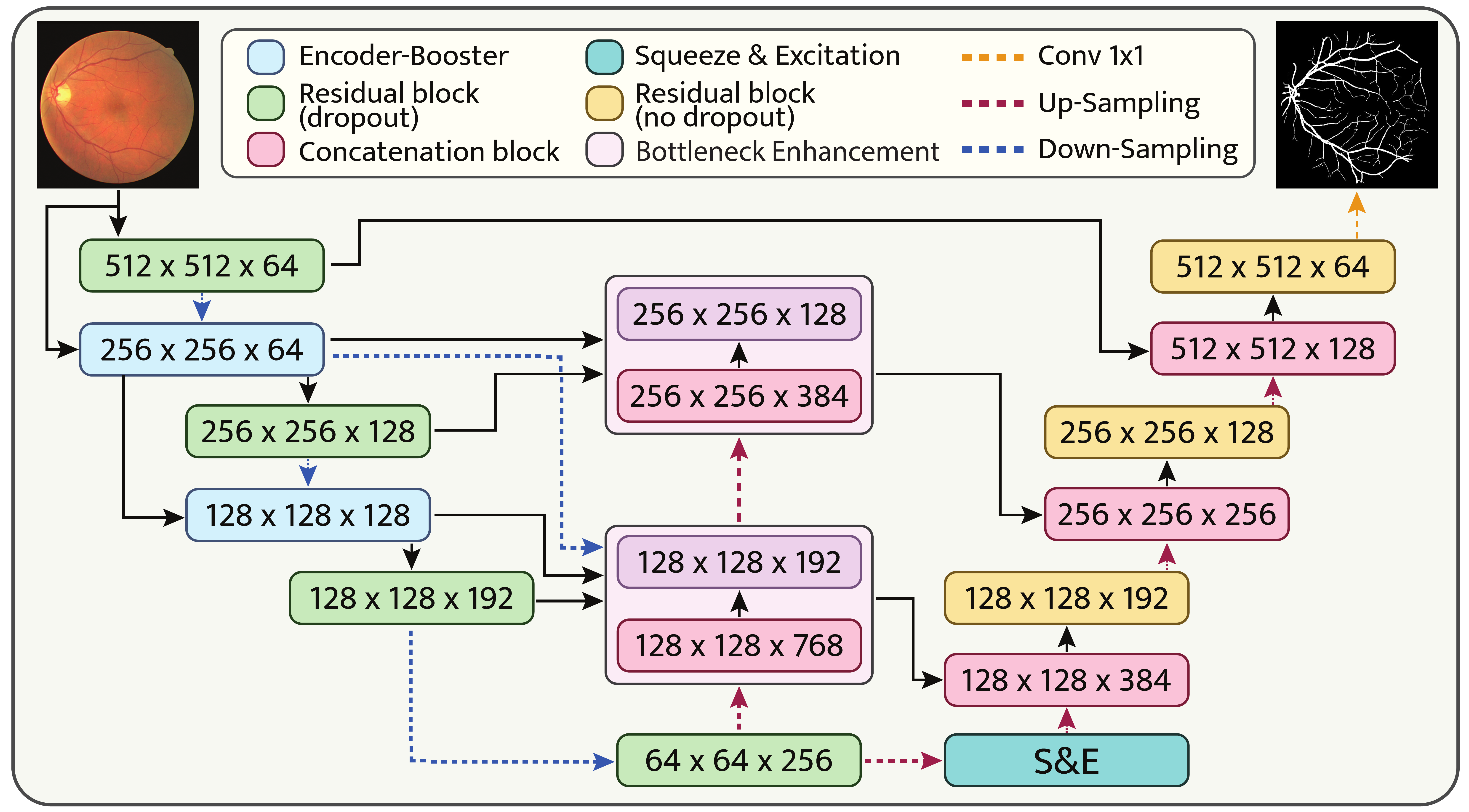}
  \caption{Main architecture}
  \label{fig:architecture}
\end{subfigure}%
\begin{subfigure}{.5\textwidth}
  \centering
  \includegraphics[height=0.55\linewidth]{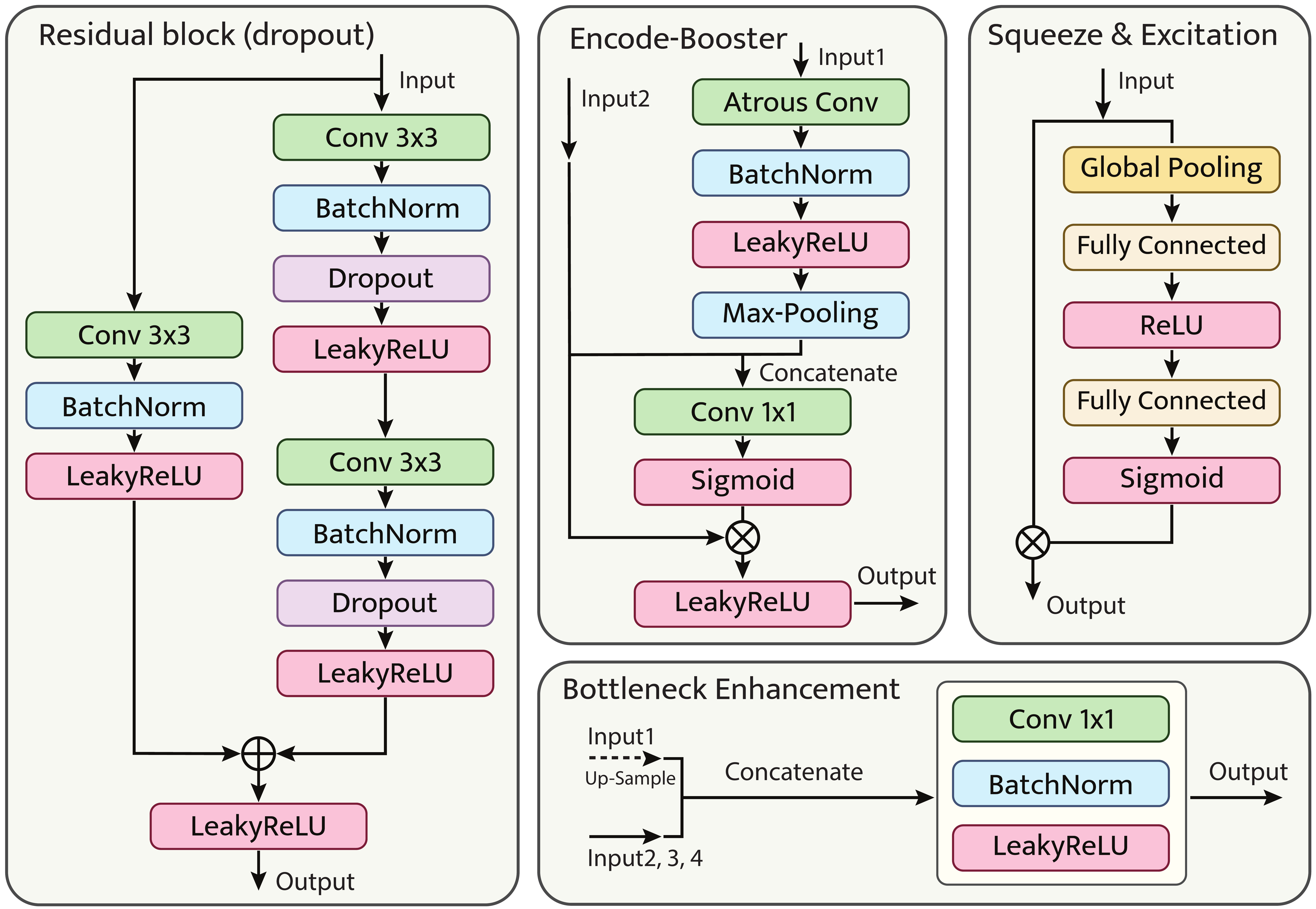}
  \caption{Building blocks of the main architecture}
  \label{fig:all_block}
\end{subfigure}
\caption{The proposed full-scale neural network structure.}
\label{fig:aaa}
\end{figure*}
\paragraph{Transformers}
HT-Net \cite{hu2022ht} proposed a future fusion block and a future refinement block to solve the problem of extracting different sizes of vessels. To capture long-range dependencies, HT-Net inserted a transformer with self-attention at the bottom of the network. Similarly, to address multi-scale information and dependencies of vascular structures, MTPA-UNet \cite{jiang2022mtpa_unet} proposed a combination of convolutional and transformative methods. It includes transformer architecture components that are integrated with an encoder-decoder architecture, demonstrating its promising performance.

\section{Methodology}

The proposed architecture, as shown in Figure \ref{fig:aaa}, is described below. 

\paragraph{Residual Block} \label{residual block}
A new residual block is used instead of the original blocks of U-Net architecture. It is responsible for multi-resolution feature representation by avoiding the vanishing of gradient descent when the depth of the network increases \cite{he2016deep}. As observed in Figure \ref{fig:all_block}, it has a dropout layer to avoid overfitting, and this dropout is included only at the encoder side and not used at the decoder side. The reason is that the decoder part is used for feature up-sampling, and having the dropout layer leads to poor performance of the model.


\paragraph{Encoder Booster}
As shown in Figure \ref{fig:architecture}, the left side is an encoder architecture responsible for the feature extraction of the retinal vessels. The primary operations of this block are the convolutional layer, batch normalization layer, max-pooling, and leaky-ReLU activation function. As the network goes deeper, the extraction of the global features will increase while the local features will start to disappear in the process. To achieve this, the skipping connections are used to map these features to the corresponding decoder side to enhance feature visibility, though this cannot fully solve the problem at hand. Notwithstanding that the global features are extracted efficiently when the networks get deeper, thin vessel features will be lost due to the resolution loss at the encoder thus making it impossible to retrieve these features in the decoder \cite{kamran2021rv}. This is caused by the repeated down-sampling operations. To allow the encoder block to extract both the local and the global features while also retaining the micro-vessel structures, the block shown in Figure \ref{fig:all_block} is introduced. This block is integrated between the consecutive encoder blocks and uses two inputs: from the previous encoder and the encoder booster (except for the first booster, which takes input from the previous encoder and the input to the model), which as a result helps to keep better local and global information than using the standalone encoder blocks. To boost the encoder feature representation ability of micro-vascular structures, atrous convolution \cite{chen2017deeplab} is used instead of the usual one to be able to control the feature resolutions.

\paragraph{Bottleneck Enhancement}
In U-shaped segmentation models \cite{ronneberger2015u}, the bottleneck block, situated between the encoder and decoder, plays a pivotal role. It diverges from conventional encoders by not using a max-pooling operation and is essential for enhancing feature representation and localization for the decoder. This block compresses the input, retaining key features for high-resolution segmentation map reconstruction.

Our research, drawing inspiration from U-Net3+ \cite{huang2020unet}, introduces a bottleneck enhancement block. This new component, devoid of the decoder-side dense skip connection to reduce feature redundancy, comprises an up-sampling element and a normalized convolutional layer. It is designed to augment the dense prediction of micro-vessel structures, a task beyond the reach of standard bottlenecks. The block's integration with the encoder and bottleneck shows promise in enhancing the up-sampling process.

Let \( E: \mathcal{X} \to \mathcal{Y} \) and \( D: \mathcal{Y} \to \mathcal{Z} \) be the encoder and decoder functions, respectively, mapping between input space \(\mathcal{X}\), feature space \(\mathcal{Y}\), and output space \(\mathcal{Z}\). The bottleneck block \( B: \mathcal{Y} \to \mathcal{Y}' \) optimizes feature representation for improved localization and reconstruction, expressed as:
\begin{equation}
   \min_{B} \; \mathcal{L}(D(B(E(x))), x') 
\end{equation}
where \( \mathcal{L} \) is a loss function, \( x \) is the input, and \( x' \) is the desired output.

\paragraph{Squeeze and Excitation Block} \label{SE}
In order to access more contextual information and capture channel-wise dependencies, squeeze and excitation operations are used, respectively \cite{hu2018squeeze}. Let $F$ be the feature map produced by a convolutional layer. The squeeze operation $S(F)$ is defined as a global average pooling operation which compresses $F$ into a channel descriptor $S(F) =\frac{1}{N} \sum_{i=1}^N F_i$ where $N$ is the number of pixels in the feature map and $F_i$ is the feature value at pixel $i$. The excitation operation is therefore the function $E(S) = \sigma(W_{2}\delta (W_{1}S))$ where $\sigma$ is the sigmoid function, $\delta$ is a ReLU function, and $W_{1},W_{2}$ are the weights of the linear layers.

\paragraph{Sigmoid Smoothing and Adaptive Threshold} \label{AT}
In retinal fundus images, there is a significant imbalance between positive and negative classes, leading to a high rate of false negatives in vessel classification. This discrepancy is evident both visually, in the comparison between the ground truth and the prediction map, and numerically, in the sensitivity scores. The model's output, without any function applied post-last layer, is processed through a sigmoid function to adjust the output features. This adjustment involves shifting 'near-negative pixels'—those that would typically be classified as negative under standard thresholding—to positive, thereby smoothing the output. Following this, an adaptive threshold algorithm is employed to determine the optimal threshold for classifying each pixel into its respective class.

The output of the model $O$ is passed through a sigmoid function $\sigma$ for smoothing $O^{\prime}=\sigma(O)$. The adaptive thresholding problem can be formulated as finding the threshold $\theta$ that minimizes the difference between the predicted class and the actual class. Let $C$ be the class label (positive or negative). The optimization problem can be defined as:
\begin{equation}
\min _\theta \sum_{i=1}^M\left|\mathds{1}_{\left\{O_i^{\prime} \geq \theta\right\}}-C_i\right|   
\end{equation}
where $M$ is the number of pixels, $O_i^{\prime}$ is the sigmoid-smoothed output for pixel $i, \mathds{1}$ is the indicator function, and $C_i$ is the true class of pixel $i$.

\RestyleAlgo{ruled}
\SetKwComment{Comment}{/* }{ */}

\begin{algorithm}[hbt!]
\caption{Adaptive Threshold Algorithm}\label{alg:adaptive_threshold}
\KwData{Sigmoided prediction matrix \( P \) of size \( H \times W \)}
\KwResult{Optimal threshold \( \theta^* \)}
\SetKwInOut{Initialization}{Initialization}
\Initialization{\( \theta \leftarrow \theta_{\text{initial}} \)}
\While{true}{
    \( \text{Thresholded} \leftarrow \mathbf{0}_{H \times W} \) \Comment*[r]{Initialize a zero matrix}
    \For{\( i \leftarrow 0 \) \KwTo \( H-1 \)}{
        \For{\( j \leftarrow 0 \) \KwTo \( W-1 \)}{
            \eIf{\( P[i][j] \geq \theta \)}{
                \( \text{Thresholded}[i][j] \leftarrow 1 \)
            }{
                \( \text{Thresholded}[i][j] \leftarrow 0 \)
            }
        }
    }
    \( \text{foreground} \leftarrow \text{count\_nonzeros}(\text{Thresholded}) \)\\
    \( \text{background} \leftarrow \text{count\_zeros}(\text{Thresholded}) \)\\
    \( \text{ratio} \leftarrow \frac{\text{background}}{\text{foreground}} \)\\
    \If{\( \left| \text{optimum} - \text{ratio} \right| \leq \epsilon \)}{
        break
    }
    \( \theta \leftarrow \theta + \Delta \theta \) \Comment*[r]{Increment threshold}
}
\textbf{return} \( \theta \)
\end{algorithm}

In binary pixel classification, the common threshold is $\theta = 0.5$, although this is not necessarily true when there is a substantial class imbalance. As a result, there is a high mis-segmentation of vessel pixels since they are few in number relative to the overall number of pixels observed in the fundus image. The suggested approach addresses this issue by tracing the pixel-level shift caused by the sigmoid smoothing used. The main point is that the background-to-vessel ratio in the retinal vessel data can be analytically approximated. This ratio is then used as an optimal point to identify the appropriate threshold for the prediction map.


\section{Data}
\begin{table}[h!]
    \caption{Dataset used in this research work}
    \label{tab:dataset}
    \begin{tabular} {p{2.7cm}p{0.5cm}p{1.5cm}p{0.5cm}p{1.5cm}}
     \toprule
     Data & Size & Resolution & FoV & Camera\\
     \midrule
     DRIVE \cite{staal2004ridge} & 40 & 565$\times$584 & 45$^\circ$ &  \tiny Canon CR5 3CCD \\
     CHASE-DB1 \cite{fraz2012ensemble} & 28 & 990$\times$960 & 30$^\circ$ &  \tiny Nidek NM-200-D \\
     STARE \cite{hoover2000locating} & 20 & 700$\times$605 & 35$^\circ$ &  \tiny TopCon TRV-50 \\
     \myRed{DCA1} \cite{cervantes2019automatic} & 134  & 300$\times$300 & -- & -- \\
     \bottomrule
    \end{tabular}
\end{table}

The DRIVE dataset is divided into 20 training images and 20 testing images. The CHASE-DB1 has 20 training images and 8 testing images. Because the STARE dataset only contains 20 images, \textit{k}-fold cross-validation is utilized, with \textit{k} set to 20. The first three datasets are independently annotated by two experts, and the one annotated by the first expert is used in this work. \myRed{The DCA1 dataset consists of a total of 134 images with their corresponding ground truth annotated by an expert cardiologist, and it was randomly split into 104 images for training and 30 images for testing.}

\section{Results and Discussion}
\subsection{Experiment Details}
 \begin{figure*}[h]
    \centering
    \includegraphics[width=1\textwidth]{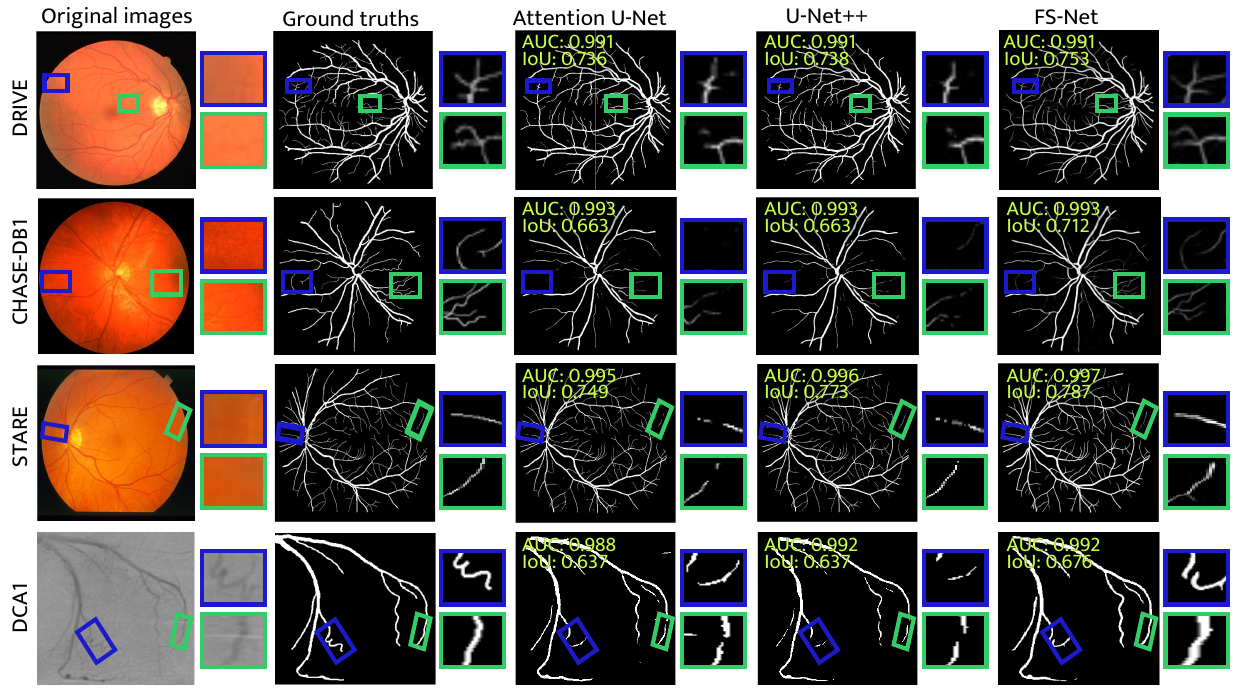}
    \caption{Comparison of prediction maps of different models with this work on the Four datasets.}
    \label{fig:prediction}
\end{figure*}

The implementation of the proposed work was carried out using the Pytorch framework. The model is trained on a single NVIDIA GeForce GTX 1080ti GPU with 12 GB. Adam optimizer with a learning rate of \texttt{1e-3} and a binary cross-entropy loss function are applied. The model was trained with a batch size of 2 for 100 epochs. To compensate for the small size of the data, several data augmentation techniques were performed, including  rotation, flipping, optical distortion, gamma correction, and equalizing histograms. DRIVE, CHASE-DB1, and STARE datasets are resized to 565$\times$565, 960$\times$960, and 605$\times$605 pixels, respectively. Then, it is converted to grayscale before feeding to the network. To ensure the best results, we conducted the validation during the training phase and saved the model with the lowest loss on the validation data for the final inference and metric evaluation on test data.

\begin{table*}[h!]
\centering
\caption{Performance Comparison on DRIVE, CHASE-DB1, STARE, and \myRed{DCA1} datasets.}
\label{tab:performance}
\begin{tabular} {p{2cm}p{3cm}p{1.1cm}p{1.1cm}p{1.1cm}p{1.1cm}p{1.1cm}p{1.1cm}p{1.1cm}}
 \toprule
Dataset & Methods & Sen. & Spe. & F1 & Acc. & AUC & IoU \\
 \midrule
\multirow{10}{*}{DRIVE}
& \myRed{Zaho et al.}\cite{zhao2015automated} & 0.742 & 0.982 & -- & 0.954 & 0.862 & -- \\
& U-Net\cite{ronneberger2015u}  & 0.7881 &	0.987 &	0.8176 & 0.9695 & 0.9874 & 0.6918\\
& Attention UNet \cite{oktay2018attention}   & 0.7999 & 0.986	& 0.82 & 0.9695 & 0.9875 & 0.6953\\
& U-Net++\cite{zhou2018unet++}  & 0.8055 & 0.9854 & 0.8208 & 0.9695 & 0.9874 & 0.6964\\
& U-Net3+\cite{huang2020unet}   & 0.7874 & 0.9869	& 0.8166 & 0.9693 & 0.987 & 0.6905\\
& \myRed{Alhussein et al.}\cite{alhussein2020unsupervised} & 0.7851 & 0.9724 & -- & 0.9559 & 0.878 & -- \\
& RV-GAN\cite{kamran2021rv}   & 0.7927 & \textbf{0.9969} & \textbf{0.8690} & \textbf{0.9790} & 0.9887 & --\\
& SGL\cite{zhou2021study}   & 0.8380 & 0.9834 & 0.8316 & 0.9705 & 0.9886 & --\\
& SA-UNet \cite{guo2021sa}   & 0.8212 & 0.9840 & 0.8263 & 0.9698 & 0.9864 & --\\
& FR-UNet \cite{liu2022full}   & 0.8356 & 0.9837 & 0.8316 & 0.9705 & \textbf{0.9889} & \textbf{0.7120}\\
& \textbf{FS-Net}  & \textbf{0.8421} & 0.9828 & 0.8314 & 0.9702 & 0.9884 & 0.7116\\
\midrule
 
\multirow{8}{*}{CHASE-DB1}
& U-Net\cite{ronneberger2015u} & 0.7968 & 0.9876 & 0.8038 & 0.9755 & 0.9891 & 0.6723\\
& Attention UNet \cite{oktay2018attention}   & 0.7505 & 0.9908 & 0.795 & 0.9756 & 0.9902 & 0.6601\\
& U-Net++\cite{zhou2018unet++}  & 0.7468 & \textbf{0.9914} & 0.7961 & 0.9759 & 0.9901 & 0.6618\\
& \myRed{Alhussein et al.}\cite{alhussein2020unsupervised} & 0.7776 & 0.9634 & -- & 0.9505 & 0.8705 & -- \\
& RV-GAN\cite{kamran2021rv}   & 0.8199 & 0.9806 & \textbf{0.8957} & 0.9697 & 0.9914 & --\\
& SGL\cite{zhou2021study}   & 0.8690 & 0.9843 & 0.8271 & \textbf{0.9771} & \textbf{0.9920} & --\\
& FR-UNet \cite{liu2022full}   & \textbf{0.8798} & 0.9814 & 0.8151 & 0.9748 & 0.9913 & \textbf{0.6882}\\
& \textbf{FS-Net}   & 0.8365 & 0.985 & 0.8144 & 0.9755 & 0.9903 & 0.683\\
\hline
 
\multirow{7}{*}{STARE}
& U-Net\cite{ronneberger2015u}  & 0.8331 & 0.9842 & 0.82 & 0.9729 & 0.9903 & 0.6972\\
& Attention UNet \cite{oktay2018attention}  & 0.7813	& \textbf{0.9865} & 0.7948 & 0.9714	& 0.9875 & 0.6673\\
& U-Net++\cite{zhou2018unet++}  & 0.8392 & 0.9839 & 0.8231 & 0.9733 & 0.9904 & 0.7009\\
& R2U-Net \cite{alom2019recurrent}   & 0.8298 & 0.9862 & \textbf{0.8475} & 0.9712 & 0.9914 &  --\\
& U-Net3+\cite{huang2020unet}   & 0.8377 & 0.9813 & 0.8086 & 0.9708 & 0.9791 & 0.6807\\
& RV-GAN\cite{kamran2021rv}   & 0.8356 & 0.9864 & 0.8323 & \textbf{0.9754} & 0.9887 & --\\
& \textbf{FS-Net}  & \textbf{0.8393} & 0.9852 & 0.8261 & 0.975 & \textbf{0.9916} & \textbf{0.7055}\\
\midrule

\multirow{5}{*}{\myRed{DCA1}}
& U-Net\cite{ronneberger2015u}  & 0.7636 & 0.9895 & 0.7802  & \textbf{0.9764} & 0.9908 & 0.6429\\
& Attention UNet \cite{oktay2018attention}   & 0.7755 & 0.9866 & 0.7687 & 0.9744 & 0.9895 & 0.6276\\
& U-Net++\cite{zhou2018unet++}  & 0.7451 & 0.9893 & 0.7661 & 0.9751 & 0.9907 & 0.625\\
& U-Net3+\cite{huang2020unet}   & 0.6996 & \textbf{0.9928} & 0.7611 & 0.9759 & 0.9372 & 0.6196 \\
& \textbf{FS-Net} & \textbf{0.8365} & 0.9844 & \textbf{0.7909 } & 0.9757 & \textbf{0.9918} & \textbf{0.6559}\\
\bottomrule

\end{tabular}
\end{table*}


\subsection{Performance Evaluation}
The performance of this work is compared with the state-of-the-art studies including U-Net \cite{ronneberger2015u}, Attention U-Net \cite{oktay2018attention}, U-Net++ \cite{zhou2018unet++}, U-Net3+ \cite{huang2020unet}, R2U-Net \cite{alom2019recurrent}, SGL \cite{zhou2021study}, RV-GAN \cite{kamran2021rv}, FR-UNet \cite{liu2022full}. We have trained and evaluated the first four U-Net versions, and for the rest of the works, the performance reported by the authors is taken for comparison.

Our work surpasses others by allowing our network architecture to handle input images of various sizes, eliminating the need for resizing or cropping that can affect model performance. This approach maintains the original image size during testing, enhancing accuracy.

We trained and evaluated U-Net variants under the same conditions as our model, incorporating Gamma correction and CLAHE \cite{jiang2019automatic} on grayscale images, significantly improving fundus image clarity. Our U-Net variants outperformed those in existing research.

A key challenge in retinal vessel segmentation is micro-vessel structure extraction, a limitation of previous studies. Our method achieved a high sensitivity of 0.8421 on the DRIVE dataset, indicating a substantial reduction in false negatives and enhanced micro-vessel accuracy. While RV-GAN shows promising F1-score, accuracy, and specificity, it has lower sensitivity, indicating a tendency to classify background pixels over vessel pixels. FR-UNet also reached state-of-the-art sensitivity and AUC, with room for improvement.

Our method demonstrates competitive performance across various metrics in the DRIVE and CHASE-DB1 datasets, as detailed in Table \ref{tab:performance}. In the STARE dataset, we achieved impressive results in Sensitivity, AUC, and IoU score. Figure \ref{fig:prediction} compares our segmentation maps against previous studies, particularly highlighting our superior extraction of microvasculature structures and thin vessel branches, areas where previous efforts introduced more false negatives.

\myRed{To demonstrate the adaptability of our model across different conditions, we tested our work on another medical segmentation dataset known as DCA1. This dataset comprises X-Ray coronary angiograms characterized by comparatively low-resolution grayscale images. Since the dataset does not come with an official training and testing split, researchers must randomly partition it for their analyses. Consequently, this lack of a standardized split makes it challenging to compare the performance of our model with other studies. To facilitate a valid comparison, we trained four different models from scratch using the same dataset configuration. Our model achieved superior performance compared to other works, as shown in \reftab{tab:performance}.}

\myRed{The proposed approach also has advantages regarding computational complexity compared to previous works. Although the model is not considered to be very lightweight, which is not our primary concern, the size of the parameters is acceptable considering the complexity of integrated architectures. The number of parameters and the floating-point operations for the U-Net architecture, its variants, and our model are presented in \reftab{tab:param-flops}.}

\begin{table}[h]
    \centering
    \caption{\myRed{Complexity comparison between different models, where the first FLOPs column corresponds to an input size of 1$\times$512$\times$512 and the second column corresponds to an input size of 1$\times$48$\times$48.}}
    \label{tab:param-flops}
    \begin{tabular}{c c c c}
    \toprule
         Methods & Params (M) & FLOPs (G) & FLOPs (G)  \\ \midrule
         U-Net\cite{ronneberger2015u} &  31.04 & 218.52 & 1.92 \\
         Attention U-Net \cite{oktay2018attention}  & 34.88 & 266.15 & 2.34 \\
         U-Net++\cite{zhou2018unet++} & 36.63 & 551.95 & 4.85 \\
         U-Net3+\cite{huang2020unet} & 26.97 & 791.56 & 6.96 \\
         \textbf{FS-Net} & \textbf{7.14} & \textbf{216.95} & \textbf{1.9} \\
         \bottomrule
    \end{tabular}
\end{table}

\subsection{Ablation Study}
In order to ascertain the effect of each component of the proposed solution, we conducted an ablation study by considering U-Net as a baseline (BL). We recall that the blocks used in the original U-Net architecture are replaced by a residual block shown in Figure \ref{fig:all_block}. As a second study, the encoder booster (EB) block presented in Figure \ref{fig:all_block} is included, and as a third study, the bottleneck enhancement (BE) block shown in Figure \ref{fig:all_block} is included. Finally, the squeeze and excitation (SE) block and adaptive threshold (AT) algorithm are included, as discussed in the methodology section. 

\begin{table*}[h!]
  \begin{threeparttable}
    \caption{Ablation study of the proposed work on DRIVE dataset.}
    \label{tab:ablation}
    \begin{tabular} {p{4.5cm}p{1.5cm}p{1.5cm}p{1.5cm}p{1.5cm}p{1.5cm}p{1.5cm}}
     \toprule
     Methods & Sen. & Spe. & F1 & Acc. & AUC & IoU \\
     \midrule
     BL & 0.7827 & \textbf{0.9887} & 0.8218 & 0.9704 & 0.9877 & 0.6981\\
     BL+EB  & 0.7894 & 0.9882 & 0.8236 & 0.9706 & 0.9878 & 0.7007\\
     BL+EB+BE  & 0.7929 & 0.9878 & 0.824 & 0.9705 & 0.988 & 0.7012\\
     BL+EB+BE+SE  & 0.7901 & 0.9886 & 0.8259 & \textbf{0.9709} & 0.9884 & 0.7038\\
     BL+EB+BE+SE+AT  & \textbf{0.8421} & 0.9828 & \textbf{0.8314} & 0.9702 & \textbf{0.9884} & \textbf{0.7116}\\
     \bottomrule
    \end{tabular}
    \begin{tablenotes}
      \small
      \item \textbf{Note:} Here BL stands for baseline network, EB for encoder booster, BE for bottleneck enhancement, SE for squeeze and excitation, and AT for adaptive threshold.
    \end{tablenotes}
  \end{threeparttable}
\end{table*}

\begin{table*}
\centering
\begin{threeparttable}
\caption{Performance of cross-training evaluation.}
\label{tab:cross training}
\begin{tabular} {p{1.81cm}p{1.81cm}p{2.5cm}p{1.52cm}p{1.52cm}p{1.52cm}p{1.52cm}}
 \toprule
Train & Test & Methods & Sen. & Spe. & Acc. & AUC\\
 \midrule
\multirow{4}{*}{DRIVE} & \multirow{4}{*}{STARE}
  &  HAnet \cite{wang2020hard} & 0.8187 & 0.9699 & 0.9543 & 0.9648 \\
  && RE-GAN \cite{zhou2021refined} & 0.8334 & \textbf{0.9764} & 0.9613 & 0.9718 \\
  && MFI-Net \cite{ye2022mfi} & 0.7805 & 0.9741 & 0.9550 & 0.9747 \\
  && \textbf{FS-Net} & \textbf{0.892} & 0.9712 & \textbf{0.9651} & \textbf{0.9871} \\
 \hline
\multirow{4}{*}{STARE} & \multirow{4}{*} {DRIVE}
  &  HAnet \cite{wang2020hard} & 0.7140 & 0.9879 & 0.9530 & 0.9758 \\
  && RE-GAN \cite{zhou2021refined} & \textbf{0.7412} & 0.9830 & 0.9519 & 0.9643 \\
  && MFI-Net \cite{ye2022mfi} & 0.7313 & 0.9867 & 0.9538 & 0.9762 \\
  && \textbf{FS-Net} & 0.7402 & \textbf{0.9889} & \textbf{0.967} & \textbf{0.9793} \\ 
  \bottomrule
\end{tabular}
\begin{tablenotes}
  \small
  \item \textbf{Note:} Here Sen. stands for Sensitivity, Spe. for Specificity, and Acc. for Accuracy.
\end{tablenotes}
\end{threeparttable}
\end{table*}

A sigmoid smoothing function is used in all architecture compositions included in the ablation investigation to provide a fair comparison. The baseline is good for segmenting background pixels as it has the highest specificity. It is also worth noting that replacing the blocks with the residual design presented in Section \ref{residual block} showed improved results than the original U-Net architecture for all metrics except for sensitivity, as demonstrated in Table \ref{tab:performance}. The inclusion of the encoder booster and bottleneck enhancement module resulted in a performance increase in most metrics, thus demonstrating that these components are critical components of the architecture. The squeeze and excitement blocks also contributed to the highest accuracy and AUC. It is worth noting that the thin vessels are recognized at this stage, resulting in a high AUC, though the sensitivity remains poor. This was addressed by implementing an adaptive threshold strategy. The adaptive algorithm drastically increased the sensitivity score. As a result, we were able to achieve a high F1-score and IoU, which was made possible by the algorithm provided in Section \ref{AT}.

\subsection{Cross Training Evaluation}
The ultimate goal of any deep learning solution is to create a system that is capable of high generalization. The situation is different in the case of retinal vascular segmentation. There are three frequently used datasets, namely DRIVE, CHASE-DB1, and STARE, and all of the works presented so far have been dedicated to training and testing in a single dataset, despite the fact that all of these datasets are retinal datasets. However, to take this study to the next level, it must be able to generalize well regardless of dataset type, as in the case of real-world scenarios. There have been a number of studies on cross-training evaluation: training in one dataset and testing in another \cite{wang2020hard, zhou2021refined, ye2022mfi}. In this work, we ran cross-training analyses on the DRIVE and STARE datasets and obtained impressive results when compared to previous studies. When trained on the DRIVE dataset and tested on the STARE dataset, our model exhibited very high sensitivity, accuracy, and AUC, owing to its superior accuracy in segmenting multi-scale vascular structures. When the model is trained on STARE and evaluated on DRIVE, it achieves better results in all metrics except sensitivity. This is because the STARE dataset does not have as many small-scale and branch vessels as DRIVE, but it is still just 0.001 behind RE-GAN.


\section{Conclusion}
In this research, we contributed to the main challenging problem of microvascular extraction in retinal vessel segmentation. The proposed solutions have different components, such as encoder boosters to keep track of information extracted by the encoder architecture, bottleneck enhancement blocks where full-scale features are learned better, squeeze and excitation blocks where feature representation of the network is improved by capturing channel dependencies, sigmoid smoothing, and adaptive threshold algorithms where microvascular segments are extracted at a high rate. The significance of all these components was evaluated by conducting an ablation study, and the results obtained demonstrated their effectiveness for the application at hand. Further, the proposed solution was also passed through cross-training evaluation in order to ascertain how well our model, when trained on one dataset, can generalize when it is tested on another dataset, and as a result, good performance was recorded. 
\myRed{
Future work will focus on addressing the three-fold generalization challenges that emerged during the course of our research: visual and degradation style shifts, diagnostic pattern diversity, and data imbalance. These issues are prevalent in complex real-world scenarios, such as microvascular extraction in retinal vessel segmentation. To tackle these challenges, we aim to develop a novel unified and generalizable framework that can effectively adapt to diverse clinical environments and data variations. Further enhancements of our model will integrate advanced techniques for domain adaptation, robust feature learning, and new data augmentation strategies to mitigate the impact of these generalization issues. 
}
\section{Acknowledgements}
Part of this work was supported by the Skolkovo Institute of Science and Technology - University of Sharjah Joint Projects: Artificial Intelligence for Life, project "Towards an Explainable Diabetic Retinopathy Grading Model".

\section*{Declaration of Competing Interest.}
The authors declare that they have no known competing financial interests or personal relationships that could have appeared to influence the work reported in this paper. 





 \bibliographystyle{elsarticle-num}

 \bibliography{Bibliography.bib}

\end{document}